\documentclass[reprint,amsmath,amssymb,pra]{revtex4-2}

\usepackage{graphicx}
\usepackage{epstopdf}
\usepackage{dcolumn}
\usepackage{bm}
\usepackage{layouts}
\usepackage{ulem}
\usepackage{amsmath}
\usepackage{hyperref}
\hypersetup{
	colorlinks   = true, 
	urlcolor     = blue, 
	linkcolor    = blue, 
	citecolor   = blue 
}
\usepackage[dvipsnames]{xcolor}
\usepackage{float}
\usepackage{siunitx}

\newcommand{\blue}[1]{\textcolor{blue}{#1}}

\begin{document}

\title{All-optical production of Bose-Einstein condensates with 2~Hz repetition rate}%


\author{Mareike Hetzel$^{1,2}$} 
\email[Author to whom correspondence should be addressed. ]{mareike.hetzel@dlr.de}
\author{Martin Quensen$^{1,2}$}
\author{Jan Simon Haase$^1$}
\author{Carsten Klempt$^{1,2}$}

\affiliation{$^1$Institut für Quantenoptik, Leibniz Universit\"at Hannover, Welfengarten 1, D-30167 Hannover, Germany  \\ $^2$Deutsches Zentrum f\"ur Luft- und Raumfahrt e.V. (DLR), Institut f\"ur Satellitengeod\"asie und Inertialsensorik (DLR-SI), Callinstraße 30b, D-30167 Hannover, Germany}

\date{\today}

\begin{abstract}
Bose-Einstein condensates (BECs) of neutral atoms constitute an important quantum system for fundamental research and precision metrology.
Many applications require short preparation times of BECs, for example, for optimized data acquisition rates in scientific applications, and reduced dead times and improved bandwidths for atomic quantum sensors. 
Here, we report on the generation of rubidium BECs with a repetition rate of more than $\SI{2}{Hz}$.
The system relies on forced evaporation in a dynamically adjusted optical potential, which is created by the spatial modulation of laser beams.
Our system provides a versatile source of the ubiquitous rubidium BECs, and promotes their exploitation for high-precision atom interferometers.
\end{abstract}

\maketitle

Bose-Einstein condensates  (BECs) of neutral atoms have become important systems for an entire field of quantum physics research.
They also start to become important for applications, mainly in the field of precision metrology, with atom interferometry being a prime example~\cite{Kasevich1991, Kasevich1992, Geiger2020}.
A BEC source is beneficial for atom interferometer due to the BEC's well controlled spatial mode and the achievable low expansion velocities~\cite{Debs2011}.
For such applications, the preparation time of BECs presents an important figure of merit.
The preparation time defines the dead time of the sensor and therefore influences bandwidth and noise sensitivity~\cite{Dutta2016}.
For scientific applications, short preparation times are important as they often dominate the data acquisition rate.

Besides few exceptions~\cite{Urvoy2019,Stellmer2013a}, BECs are typically realized by forced evaporative cooling in magnetic or optical potentials.
Our previous work~\cite{Puer2023} gives an overview of atom numbers and repetition rates obtained with atom-chip magnetic traps, macroscopic magnetic traps, all-optical traps, and combinations of magnetic and optical traps. 
The current state of the art is set by BEC preparation times of $4\times10^4$ Rb atoms within $\SI{850}{ms}$ on an atom chip~\cite{Rudolph2015}, $2\times10^4$ Er atoms within $\SI{700}{ms}$ in an optical trap~\cite{Phelps2020}, and $3\times10^3$ Rb atoms within $\SI{575}{ms}$ by direct laser cooling~\cite{Vendeiro2022}.

\begin{figure}[ht]
	\includegraphics[width=0.5\textwidth]{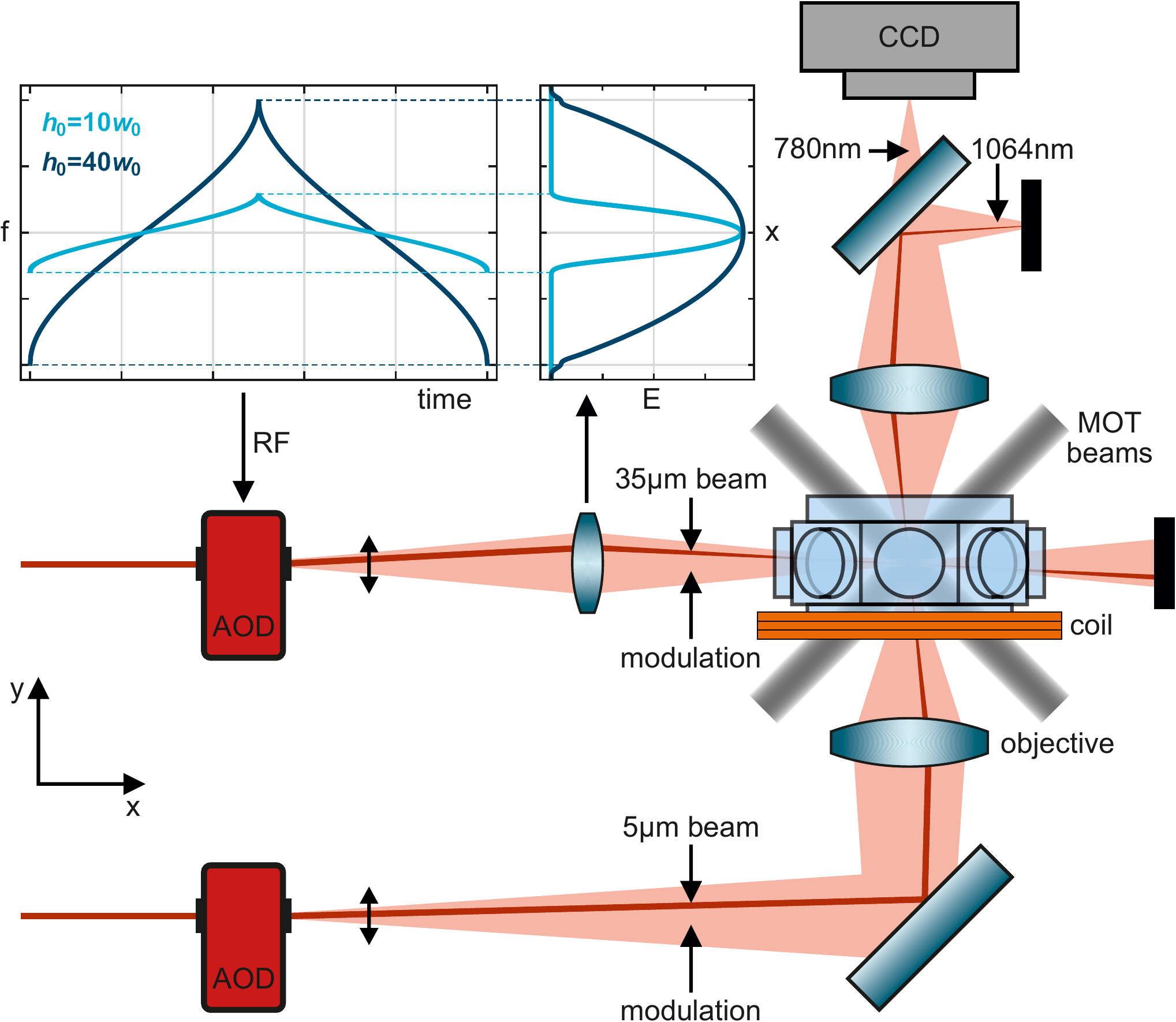}
	\caption{\label{fig:configuration} Setup for the generation of time-averaged optical potentials.
    The central evacuated octagon-shaped glass cell enables the operation of a three-dimensional magneto-optical trap (grey laser beams).
    Two laser beams (red beams) pass acousto-optical deflectors (AODs) and are focused onto the atomic ensembles to form an adjustable optical dipole trap.
    The application of modulated radio frequencies (RF, upper left inset) to the AODs shapes time-averaged spatially broadened light fields (light red beams) and corresponding adjustable harmonic optical potentials (upper right inset).
    As long as the painting stroke $h_0$ remains larger than the beam waist $w_0$ at the position of the atoms, the trapping potential's depth and width can be set independently by a suitable choice of $h_0$ and the beam power.
    A single coil creates a magnetic field gradient that is employed for spin distillation.
    The generated atomic ensemble is detected by fluorescence imaging on a CCD camera, where the fluorescence light is separated from the dipole trapping light by a dichroic mirror.}
 
\end{figure}

The all-optical generation of BECs has characteristic advantages by itself.
It does not require the operation of large coils with high currents and corresponding power and cooling considerations.
It also avoids placing objects close to the atoms inside the vacuum systems, as is required for atom-chip experiments.
The latter poses significant restrictions for atom interferometry experiments and fundamental research, where a large, unrestricted optical access to the generated BEC is highly desirable.

A fundamental problem for the all-optical generation of BECs are the conflicting requirements for initial trapping and fast evaporation.
The initial trapping of atoms from a magneto-optical trap requires large trapping volumes, while fast evaporation requires small trapping volumes to enhance the collision rate~\cite{Ketterle_1996}.
It is possible to alter the spatial mode of the trapping beams during the evaporation sequence, which has been demonstrated for example by movable optics~\cite{Kinoshita_2005}. 
Ref.~\cite{Roy2016} has pioneered the application of time-averaged (painted) potentials, where small-diameter trapping beams are spatially modulated with sufficiently high frequency to generate mean potentials whose size and shape can be altered at will.
For potassium there are approaches using Fesh\-bach resonances to lower the evaporation time of BEC generation to 170~ms~\cite{Herbst2024}.

In this letter, we report on the all-optical generation of a Rb BEC in adjustable time-averaged potentials.
We show that the quick adjustment of the potential's spatial size and depth enables short evaporation times of $\SI{275}{ms}$.
Together with a quick loading of the magneto-optical trap (MOT) of $\SI{120}{ms}$ and sub-Doppler laser cooling for $\SI{31}{ms}$, the system allows for the generation of BECs every $\SI{486}{ms}$ in continuous operation.
This includes a $\SI{50}{ms}$ final ramp-up of the potential and the $\SI{10}{ms}$ long detection sequence.
To find a minimal BEC generation time in the large parameter space, we have performed an automatic genetic optimization~\cite{Geisel2013}, yielding a stable optimum for \blue{43} parameters.

Our experimental apparatus has been described in Ref.~\cite{Puer2023} in which we demonstrated the BEC generation in a hybrid scheme of a magnetic quadrupole trap and an optical dipole trap (ODT). 
In that implementation, the effective cycle time of $\SI{5.3}{s}$ was limited by the evaporation speed in the magnetic trap and by the heating of the coils generating the magnetic field gradient. 
The experimental setup provides a three-dimensional magneto-optical trap (3D-MOT) which is loaded by a two-dimensional MOT.
The 3D-MOT has an initial loading rate of $2.4\times10^{10}$~atoms/s, and captures $4\times 10^9$~atoms within $\SI{200}{ms}$.
After MOT loading, an optical molasses cools the atoms to $\SI{18}{\micro\kelvin}$.
At the end of the molasses, the atoms are optically pumped to the $F=1$ level without preferential occupation of specific Zeeman levels.

A $\SI{55}{W}$ Coherent Mephisto MOPA laser with a wavelength of $\SI{1064}{nm}$ is split into two beams and generates a crossed-beam optical dipole trap with waists of $35$ and $\SI{5}{\micro\metre}$.
Both beams independently pass acousto-optical deflectors (AODs), which serve as control elements for an active intensity stabilization and the spatial modulation of the beams.

To account for the mode mismatch between the waists of the ODT and the laser-cooled atomic ensemble, both ODT beams are spatially modulated in horizontal direction resulting in an effectively increased trap volume, illustrated in Fig.~\ref{fig:configuration}.
The instantaneous frequency of the signal that drives the AODs is changed over time such that the deflection angle changes. 
The following optical system translates this changing angle into a horizontal translation of the ODT beams at the position of the atoms. 
The amplitude of this movement, namely the painting stroke \blue{$h_0$}, is given by the amplitude of the frequency modulation $f_s$ centered at the AOD's central frequency $f_c$ resulting in a frequency spectrum $f \in [f_c - f_s,f_c + f_s]$. 
The initial painting stroke should be matched to the size of the atomic ensemble. 
By varying the speed of the beams' movement over its spatial range, arbitrary potential shapes can be created, including harmonic potentials.
We chose a software-defined radiofrequency source (Ettus Research USRP X310) to generate the modulated frequency signals.

\begin{figure}[t]

    \includegraphics[width=0.5\textwidth]{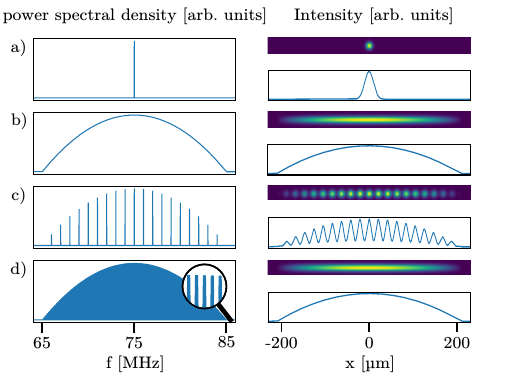}
	\caption{\label{fig:paintingeffects}
   Simulation of the effect of the painting frequency. The radiofrequency spectrum (left column), applied to the AOD, generates a time-averaged optical intensity shown in one and two dimensions (right column). (a) Without painting, a single radio frequency generates an optical potential with a Gaussian width of $\SI{5}{\micro\metre}$. (b) Low painting frequencies lead to a continuous frequency spectrum, but will result in heating as the atoms are dragged along the moving beam. (c) At a painting frequency of $\SI{1}{MHz}$, the spectrum contains sidebands that are resolved in the optical potential as discrete trapping wells. (d) At our chosen painting frequency of $\SI{100}{kHz}$, the sidebands are sufficiently dense to generate a smooth effective potential.} 
\end{figure}

An important parameter is the repetition rate of the beams' movement, namely the painting frequency $f_p$.
In our experiments, the painting frequencies $f_p$ of our beams are set to $100$ and $\SI{90}{kHz}$, respectively. 
The painting frequency has to be large compared to the trapping frequencies of the ODT $f_{x,y,z}$, such that the atoms experience a time-averaged potential. 
An upper limit is given by the Fourier spectrum of the resulting radiofrequency signal.
As illustrated in Fig.~\ref{fig:paintingeffects}, high painting frequencies become visible as resolved sidebands which lead to a fragmentation of the atomic ensemble.
At our choice of painting frequencies, the sideband structure is small compared to the waists of the beams, such that a smooth potential profile is obtained.

\begin{figure*}[t]
    \includegraphics[width=1.0\textwidth]{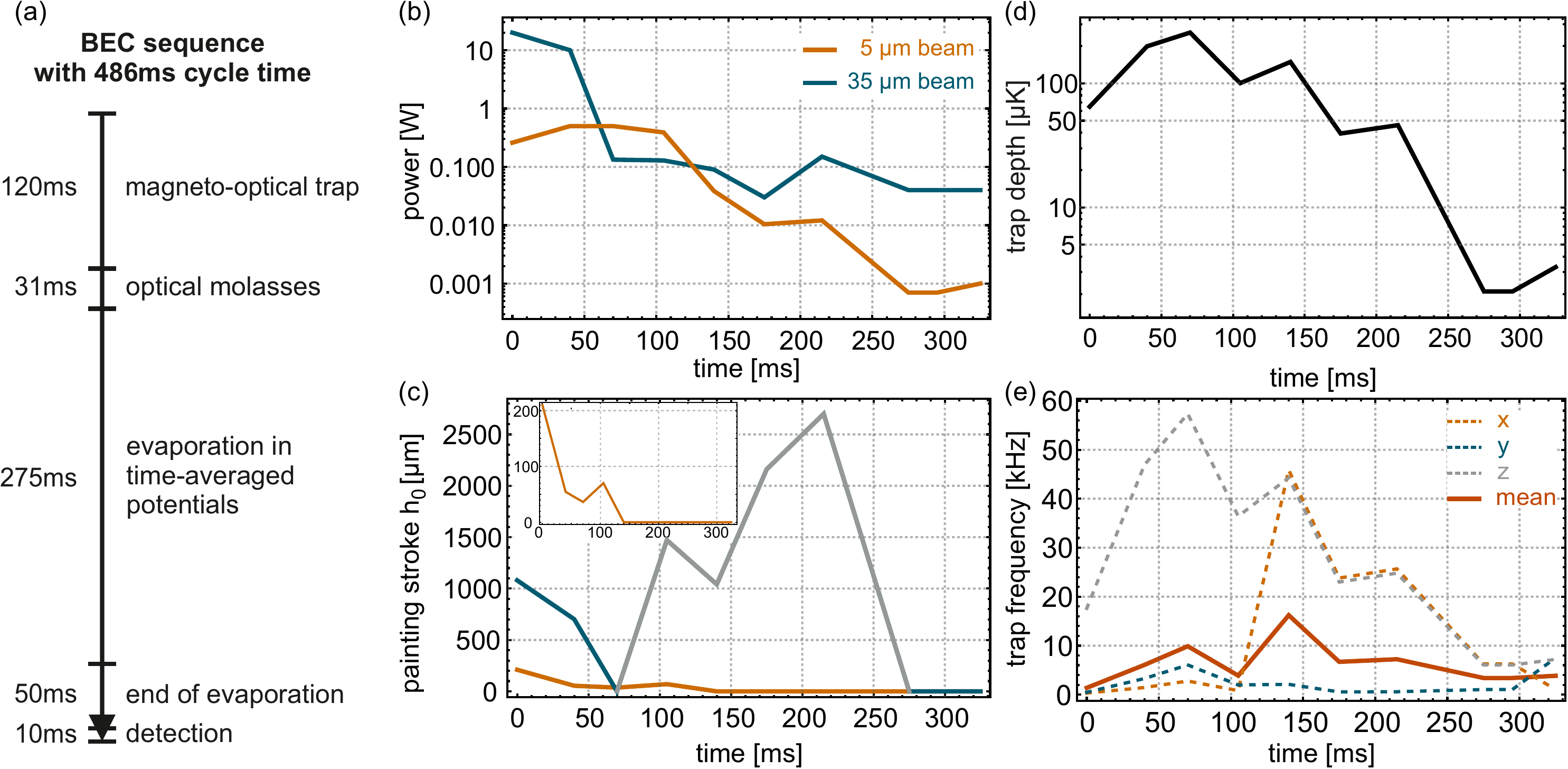}
	\caption{\label{fig:DE}
 Experimental sequence and trap parameters of the final BEC sequence. 
 (a) In the final sequence, the creation of a BEC takes a total of $\SI{486}{ms}$, $\SI{275}{ms}$ of which are for evaporation in the ODT. 
 Among the optimized parameters, the most important ones are the beam powers (b) and painting strokes (c) over the course of the ODT phase. 
 From those, the optical dipole trap depths (d) and frequencies (e) are calculated. 
 The solid grey line in (c) shows the \SI{35}{\micro\metre} beam's painting stroke during the phase in which it can be largely varied without significant impact.
 }
\end{figure*}

After ODT loading, the atoms are distributed in all Zeeman levels of the $F=1$ level. 
For a more efficient choice of a single Zeeman level, we implemented a spin distillation scheme during the dipole trap evaporation \cite{Couvert2008}.
A magnetic field gradient of $\SI{67}{G/cm}$ generated by a single coil reduces the trap depth of the $m_F=\pm 1$ levels, which are preferably removed from the trap resulting in an accumulation of atoms in the $m_F=0$ level. 

In this configuration, we created a first BEC of $3 \times 10^4$ atoms in $1~$s of evaporation time by manual optimization of all experimental parameters to check for the sanity of the setup and its alignment. 
However, the large number of partially coupled parameters makes the optimization challenging. 
The free choice of painting stroke and laser power allows for a decoupling of trap depth and frequency, but additionally complicates the optimization process.
Therefore, we applied an evolutionary optimization algorithm called Differential Evolution~\cite{Geisel2013}.
We performed the optimization in two steps.
First, we addressed the optical molasses and the loading of the ODT.
Here, $14$ parameters in total were optimized, namely the power and detuning of the cooling, repumping and optical pumping light, the duration of the molasses and the initial settings of the ODT beams.
Secondly, we improved the evaporation in the ODT and evaluated the best settings for the power and painting stroke of both ODT beams and the distillation field gradient in 6 ramps, a total of $29$ parameters. 
A list of the optimization parameters and their significance can be found in the supplemental material~\cite{supplements}.
The duration of each evaporation ramp has been set to $\SI{60}{ms}$ to facilitate the finding of stable parameters. 
The experimental sequence during the second optimization and the final evaporation parameters are illustrated in Fig.~\ref{fig:DE}. 
For both optimization steps, the atom number in the dipole trap at a fixed end configuration was set as an optimization goal. 
The end configuration was chosen to ensure that the optimization compares atom numbers at similar low temperatures.

The result of the optimization is as follows.
The initial trapping of atoms is mainly achieved by the $\SI{35}{\micro\metre}$-beam at a power of $\SI{20}{W}$ and a painting stroke of $\SI{1.1}{mm}$. 
Within two steps, the painting stroke is reduced to $0$ to compress the atomic ensemble and the power is reduced, simultaneously.
At this stage, the $\SI{5}{\micro\metre}$-beam has a negligible effect. 
However, as evaporation progresses and atoms accumulate in the combined potential, the $\SI{5}{\micro\metre}$-beam becomes more influential.
Its initial painting stroke of $\SI{210}{\micro\metre}$ and maximum power of $\SI{0.5}{W}$ are reduced to $\SI{0}{\micro\metre}$ and $\SI{1}{\milli\watt}$ at the end of the evaporation.
After the $\SI{35}{\micro\metre}$-beam's painting stroke reaches $0$, the $\SI{5}{\micro\metre}$-beam dominates the evaporation dynamics. 
Consequently, the $\SI{35}{\micro\metre}$-beam's settings can be varied over a wide range without significantly impacting the atoms (shown in grey in Fig.~\ref{fig:DE} (c)), except during the final stages of evaporation. 

With the obtained ODT parameters, we computed the trap depth and trapping frequencies throughout the evaporation. 
Here, the key to the fast evaporation becomes visible. 
Our setup combines a large initial trapping volume while maintaining a large mean trapping frequency throughout the evaporation. 
This is achieved by painting our tight focus horizontally, but maintaining a tight confinement in vertical direction which allows for a fast rethermalization. 

While the optimization was performed in a sequence with $\SI{3.6}{s}$ of cycle time, most of the time is consumed by $\SI{1}{s}$ of MOT loading and $\SI{2}{s}$ of image readout and processing.
To receive a BEC creation sequence with minimal total cycle time, we reduce the MOT loading to $\SI{120}{ms}$ and the duration of each ramp to values between $30$ and $\SI{40}{ms}$, optimizing the atomic flux.
At the end of the evaporation, the ODT beam powers are left constant for $\SI{20}{ms}$ and then increased slightly over another $\SI{30}{ms}$ to achieve long lifetimes.
In total, the evaporation takes $\SI{275}{ms}$.
For detection, a region of interest is defined to reduce the image readout duration, which is performed in parallel to the subsequent cycle.
Within $\SI{486}{ms}$, we create and detect BECs of $4 \times 10^4$ atoms \blue{with no discernable thermal fraction}.
\blue{We estimate \SI{350}{nK} as an upper limit of the BEC temperature.}
To characterize the stability of the system, two consecutive measurement series have been taken, each for $49$~s with $101$ experimental cycles with a break of $90$~s inbetween. 
The mean atom numbers and their standard deviation are $40{,}200 \pm 4{,}400$ and $40{,}700 \pm 3{,}900$, respectively, corresponding to an atom number stability of $10 \%$, without any signs of additional instability due to the high repetition rate.

To our knowledge, this presents the fastest creation of a BEC to date. 
The atom number could be significantly increased by improving the atomic flux of the 2D-MOT.
The flux was limited because we only operated the Rb dispenser, our main source of atoms, conservatively, because it harms our number-resolving detection~\cite{Puer2023} required for further experiments.
The method could be improved along the following lines.
(i) The discretization presented in Fig.~\ref{fig:paintingeffects} could be used to generate high phase-space densities when combined with the optical molasses phase.
(ii) The required painting frequency spectra can be generated by much more sophisticated time-domain signals with a possible improvement on the resulting frequencies.
(iii) Painting was only applied in the two horizontal directions which leads to a small transfer efficiency of the available atoms of the molasses phase. 
An extension to the third dimension is complicated by the gravitational potential.
With increased optical power, discretized trapping potentials could be spread along the third dimension to collect the atoms, and subsequently unify them during the evaporation.

\begin{acknowledgments}
We would like to thank Henning Albers, Knut Stolzenberg, Sebastian Bode, Dorothee Tell and Dennis Schlippert for the collaborative development of the programming code of the dipole trap potentials and the control of the software-defined radio. 
Aditionally, we want to thank Alexander Herbst for valuable discussions on time-averaged potentials. 

We acknowledge financial support from the Deutsche Forschungsgemeinschaft (DFG, German Research Foundation) - Project-ID 274200144 - SFB 1227 DQ-mat within project B01 and Germany’s Excellence Strategy - EXC-2123 QuantumFrontiers-Project-ID 390837967. 
M.Q. acknowledges support from the Hannover School for Nanotechnology (HSN).
J.S.H. acknowledges funding by the German Space Agency DLR with funds provided by the Federal Ministry of Economics and Technology (BMWi) under grant number 50WM2174.
\end{acknowledgments}

\bibliography{main,bib2}

\end{document}